\documentclass[aps,showpacs, nofootinbib]{revtex4}

\newcommand{\be}{\begin{equation}}
\newcommand{\ee}{\end{equation}}
\newcommand{\ben}{\begin{eqnarray}}
\newcommand{\een}{\end{eqnarray}}

\newcommand{\la}{{\lambda}}

\newcommand{\cL}{{\cal L}}

\newcommand{\p}{\partial}
\newcommand{\na}{\nabla}

\newcommand{\trho}{{\tilde \rho}}

\newcommand{\hg}{{\hat g}}

\newcommand{\hna}{{\hat {\nabla}}}

\newcommand{\tna}{{\tilde {\nabla}}}

\newcommand{\tg}{\tilde g}

\newcommand{\tR}{\tilde R}
\newcommand{\tG}{\tilde G}
\newcommand{\tH}{\tilde H}

\newcommand{\hR}{\hat R}

\newcommand{\ep}{\epsilon}

\newcommand{\ga}{\gamma}

\newcommand{\hSi}{\hat \Sigma}


\begin{document}

\title{Uniqueness of charged static asymptotically flat black holes in dynamical Chern-Simons gravity}

\author{Marek Rogatko}
\affiliation{Institute of Physics \protect \\
Maria Curie-Sklodowska University \protect \\
20-031 Lublin, pl.~Marii Curie-Sklodowskiej 1, Poland \protect \\
marek.rogat@poczta.umcs.lublin.pl \protect \\
rogat@kft.umcs.lublin.pl}

\date{\today}

\begin{abstract}
Making use of the conformal positive energy theorem 
we prove the uniqueness of four-dimensional static {\it electrically} charged black holes being the solution
of Chern-Simons modified gravity equations of motion.
We assume that black hole spacetime 
contains an asymptotically flat spacelike hypersurface with compact interior and
non-degenerate components of the event horizon.
\end{abstract}

\maketitle

\section{Introduction}
Gravitational collapse and emergence of black holes is one of the most essential research problem
of general relativity and its generalizations.
The problem of classification of non-singular black hole solutions was first discussed by
Israel \cite{isr},
M\"uller zum Hagen {\it et al.} \cite{mil73} and Robinson \cite{rob77}, while
the most complete results were proposed in Refs.\cite{bun87,ru,ma1,he1,he93}.
The classification of static vacuum black hole solutions was finished 
in \cite{chr99a}, where
the condition of non-degeneracy of the 
event horizon was removed. As far as the Einstein-Maxwell (EM) black holes was concerned
it was proved that for the static
electro-vacuum black holes all degenerate components of 
the event horizon should have charges of the same signs \cite{chr99b}.
\par
The 
construction of the uniqueness black hole theorem for stationary axisymmetric
spacetime turned out to be far more complicated task \cite{stat}. However, the complete proof was presented by
Mazur \cite{maz} and Bunting \cite{bun}
(see for a review of the uniqueness of black hole
solutions story see \cite{book} and references therein).

A different issue, related to the problem of
gravitational collapse in generalization of Einstein theory to higher dimensions and
emergence of higher dimensional black objects (like black rings, black Saturns) and 
multidimensional black holes
was widely studied. The complete classification of $n$-dimensional 
charged black holes both with non-degenerate and 
degenerate component of the event horizon was proposed in Refs.\cite{nd}, while
partial results for the very nontrivial case of $n$-dimensional rotating black hole uniqueness theorem were 
provided in \cite{nrot}. 
The problem of the behaviour of matter fields in the spacetime of higher dimensional 
black hole was studied in Ref.\cite{rog12}.
\par
Due to the attempts of building a consistent quantum gravity theory there was also
resurgence of works treating the mathematical aspects of the low-energy string theory
black holes.
These researches comprise also the case of the low-energy limit 
of the string theory, like dilaton gravity, Einstein-Maxwell-axion-dilaton (EMAD)-gravity and 
supergravities theories \cite{sugra}. On the other hand,
the strictly stationary static vacuum spacetimes in Einstein-Gauss-Bonnet theory were
discussed in \cite{shi13a}
\par
Black holes and 
their properties as key ingredients of the AdS/CFT attitude \cite{adscft} to superconductivity
also acquire great attention. Questions about possible matter configurations in AdS spacetime 
arise naturally during aforementioned researches. In Ref.\cite{shi12}
it was revealed that strictly stationary AdS spacetime could not allow for the existence of nontrivial
configurations of complex scalar fields or form fields. The generalization of the aforementioned problem,
i.e., strictly stationarity of spacetimes with complex scalar fields
in EMAD-gravity with negative cosmological constant was given in \cite{bak13}.
\par
The Chern-Simons modified gravity (CS modified gravity), where the Einstein action is modified by the 
addition of parity violating Pontryagin
term \cite{jac03} has its roots in particle physics. Namely, the 
imbalance between left-handed and right-handed fermions induced 
gravitational anomaly in fermion number current, proportional to the aforementioned Pontryagin term \cite{alv84}.
It also emerges in string theory as an anomaly-canceling term in Green-Schwarz mechanism \cite{gre87}. Moreover
CS-gravity was elaborated in the context of cosmology, gravitational waves, Lorentz invariance \cite{ale09}
(see also references therein). 
In Ref. \cite{shi13} it was revealed that a static asymptotically flat black hole
solution is unique to be Schwarzschild spacetime in CS modified gravity.
\par
Motivated by the aforementioned problems we shall consider the problem of the uniqueness static 
asymptotically flat black holes in CS modified gravity with $U(1)$-gauge field. The basic idea in our treatment
of the problem in question will be to implement the conformal positive energy theorem \cite{sim99}.
\\
The paper is organized as follows. In Sec.II we review some basic facts concerning with
dynamical CS modified gravity. Then, applying the conformal positive
energy theorem we perform the uniqueness proof of static
asymptotically flat electrically charged black hole in CS modified gravity.

\section{system}
We commence this section with the action of the CS modified gravity with matter fields
provided by the action
\ben
I &=& \kappa~\int d^4 x \sqrt{-{g}}~R 
+ {\alpha \over 4}~\int d^4 x \sqrt{-{g}}~\theta~\ast R_{\alpha \beta \ga \delta}~R^{\alpha \beta \ga \delta}
- {\beta \over 2}~\int d^4 x \sqrt{-{g}}~\na_\alpha \theta \na^\alpha \theta \\ \nonumber 
&+& \int d^4 x \sqrt{-{g}}~\cL_{mat},
\label{act}
\een
where $\alpha, ~\beta$ are the dimensional coupling constant, while $\theta$ (CS coupling field) is scalar field
which is a function parameterizing deformation from ordinary Einstein theory. $\ast 
R_{\alpha \beta \ga \delta}~R^{\alpha \beta \ga \delta}$ is the Pontryagin density, while
$\cL_{mat}$ stands for some matter Lagrangian density which does not depend on the scalar field in question.
In what follows we assume that $\cL_{mat}$ will constitute matter Lagrangian for $U(1)$-gauge fields,
given by $\cL_{mat} = - F_{\mu \nu}F^{\mu \nu}$.
The dual to Riemannian tensor
is defined as
\be
\ast R_{\alpha \beta \ga \delta} = {1 \over 2}~\ep_{\ga \delta }{}{}^{\psi \zeta} ~R_{\alpha \beta \psi \zeta}.
\ee
The field equations obtained by variation of the action (\ref{act}) imply
\ben
R_{\mu \nu} &-& {1 \over 2}~g_{\mu \nu}R + {\alpha \over \kappa}~C_{\mu \nu}
= {1 \over \kappa} \bigg(
T_{\mu \nu} (\theta) + T_{\mu \nu}(F) \bigg),\\
\na_{\alpha} \na^{\alpha} \theta &=& {\alpha \over 4~\beta}~R_{\alpha \beta \ga \delta}~\ast 
R^{\alpha \beta \ga \delta},
\een 
where we have denoted by $C^{\alpha \beta}$ the following relation:
\be
C^{\alpha \beta} = \na_{\ga} \theta~\ep^{\ga \psi \mu (\alpha} \na_{\mu} R_{\psi}{}^{\beta)}
+ 
\na_{\ga} \na_{\delta} \theta~\ast R^{\delta (\alpha \beta) \ga}.
\ee
On the other hand, the energy momentum tensor 
$T_{\mu \nu} = - {\delta S \over \sqrt{- g}~\delta g^{\mu \nu}}$ of matter fields in question yields
\ben
T_{\alpha \beta}(\theta) &=& {\beta \over 2}~\bigg(
\na_\alpha \theta \na_\beta \theta 
- {1 \over 2}~g_{\alpha \beta}~\na_{\ga} \theta \na^{\ga} \theta \bigg), \\
T_{\alpha \beta}(F) &=& 2~F_{\alpha \ga}~F_{\beta}{}^\ga - {1 \over 2}~g_{\alpha \beta}~F_{\mu \nu} F^{\mu \nu}.
\een 
The line element of static spacetime subject to the
asymptotically timelike Killing vector field 
$k_{\alpha} = \big({\p \over \p t }\big)_{\alpha}$
and $V^{2} = - k_{\mu}k^{\mu}$ can be provided by the relation
\be
ds^2 = - V^2 dt^2 + g_{i j}dx^{i}dx^{j},
\label{met}
\ee
where $V$ and $g_{i j}$
are independent of the $t$-coordinate as the quantities
of the hypersurface $\Sigma$ of constant $t$. We assume that on the hypersurface $\Sigma$
the electromagnetic potential will be of the form $A_{0} = \psi~dt$, i.e., one deals with
electrically charged black hole.\\
In our consideration we shall take into account the asymptotically
flat spacetime. Namely, the spacetime in question will contain a data set 
$(\Sigma_{end},~ g_{ij},~ K_{ij})$ with gauge fields of $F_{\alpha \beta}$
such that $\Sigma_{end}$ is diffeomorphic to $\bf R^{3}$ minus 
a ball. The fields $(g_{ij},~ K_{ij})$ will satisfy the fall-off condition of the form
\be
\mid g_{ij} - \delta_{ij} \mid +~ r \mid \p_{a} g_{ij} \mid + \dots
+r^{m} \mid \p_{a_{1} \dots a_{m}} g_{ij} \mid + r \mid K_{ij}\mid + \dots
+r^{m} \mid \p_{a_{1} \dots a_{m}} K_{ij} \mid \le {\cal O}\bigg( {1 \over r} \bigg).
\ee
Likewise we require that in the local coordinates as above, defined $U(1)$-gauge field
fulfills the following fall-off demand:
\be
\mid F_{\alpha \beta} \mid +~ 
r \mid \p_{a} F_{\alpha \beta} \mid + \dots +
r^{m} \mid \p_{a_{1} \dots a_{m}}F_{\alpha \beta} \mid
\le {\cal O}\bigg( {1 \over r^{2}} \bigg).
\ee
In the light of these stipulation the hypersurface will be said to be asymptotically flat
if it contains an asymptotically flat end.
\par
Taking the form of static metric into account, the corresponding
equations of motion yield
\be
V~{}^{(g)}\na_{i}{}^{(g)}\na^{i} V =
{1 \over \kappa}{}^{(g)}\na_{i} \psi {}^{(g)}\na^{i} \psi,
\ee
\be
{}^{(g)}\na_{i}{}^{(g)}\na^{i} \psi = {1 \over V}{}^{(g)}\na_{i} \psi {}^{(g)}\na^{i} V,
\ee
\be
V~{}^{(g)}\na_{i} {}^{(g)}\na^{i} \theta  + {}^{(g)}\na_{i} \theta~{}^{(g)}\na^{i} V = 0,
\ee
\ben
{}^{(g)} R_{ij} &-& {{}^{(g)}\na_{i}{}^{(g)}\na_{j}V \over V}
= {1 \over \kappa}~\bigg(
{\beta \over 2}~
{}^{(g)}\na_{i}\theta {}^{(g)}\na_{j} \theta
+ g_{ij}~
{{}^{(g)}\na_{i}\psi {}^{(g)}\na^{i} \psi \over V^2}
- 2~{{}^{(g)}\na_{i}\psi {}^{(g)}\na_{j} \psi  \over V^2} \bigg).
\een
In the above relations covariant derivative with respect to the
metric tensor $g_{ij}$ is denoted by ${}^{(g)}\na$,
while ${}^{(g)}R_{ij}(g)$ is the Ricci tensor defined on 
the hypersurface $\Sigma$. Furthermore, let us suppose that for each of the quantity in question, i.e.,
$V,~\psi,~\phi$, there is a standard coordinate system in which they have usual asymptotic expansion.
\par
To proceed further, let us introduce the definitions of the crucial quantities in the 
the proof of the uniqueness. Namely, they can be written as follows:
\ben
\Phi_{1} &=& {1 \over 2} \bigg[ V + {1 \over 2~V} \bigg], \\
\Phi_{0} &=& i~\sqrt{{\beta \over 2~\kappa}}~\theta,\\
\Phi_{-1} &=& {1 \over 2} \bigg[ V - {1 \over 2~V} \bigg],
\een
and
\ben
\Psi_{1} &=& {1 \over 2} \bigg[ V + {1 \over 2~V} - \sqrt{{2 \over \kappa}}~{\psi^2
\over V} \bigg], \\
\Psi_{0} &=& {2 \over \kappa} ~{\psi \over V},\\
\Psi_{-1} &=& {1 \over 2} \bigg[ V - {1 \over2~ V} - \sqrt{{2 \over \kappa}}~{\psi^2
\over V} \bigg].
\een
It worth pointing out that defining the metric tensor $\eta_{AB} = diag(1, -1, -1)$, it can be achieved
that $\Phi_{A} \Phi^{A} = \Psi_{A} \Psi^{A} = -1$, where $A = - 1, 0, 1$.
Having in mind the conformal transformation provided by
\be
\tg_{ij} = V^{2} g_{ij},
\ee
one can introduce the symmetric tensors written in terms of $\Phi_A$
in the following form:
\be 
\tG_{ij} = \tna_{i} \Phi_{-1} \tna_{j} \Phi_{-1} -
\tna_{i} \Phi_{0} \tna_{j} \Phi_{0} -
\tna_{i} \Phi_{1} \tna_{j} \Phi_{1},
\label{g1}
\ee
and similarly for the potential $\Psi_{A}$
\be
\tH_{ij} = \tna_{i} \Psi_{-1} \tna_{j} \Psi_{-1} -
\tna_{i} \Psi_{0} \tna_{j} \Psi_{0} -
\tna_{i} \Psi_{1} \tna_{j} \Psi_{1},
\label{h1}
\ee
where by $\tna_{i}$ we have denoted the covariant derivative with respect to the metric $\tg_{ij}$.
Consequently, according to
the relations (\ref{g1}) and (\ref{h1}), the field equations
may be cast in the forms
\be
\tna^{2}\Phi_{A} = \tG_{i}{}{}^{i} \Phi_{A}, \qquad
\tna^{2} \Psi_{A} = \tH_{i}{}{}^{i} \Psi_{A}.
\label{ppff}
\ee
Just
the Ricci curvature tensor with respect to the conformally rescaled metric $\tg_{ij}$ implies
\be
\tR_{ij} = \tG_{ij} + \tH_{ij}.
\label{rr}
\ee
In general, as far as the conformal positive energy theorem is concerned,
one assumes that we have to do with two asymptotically flat Riemannian $(n-1)$-dimensional manifold
$(\Sigma^{\Phi},~ {}^{\Phi}g_{ij})$ and $(\Sigma^{\Psi},~ {}^{\Psi}g_{ij})$. Moreover the conformal transformation 
of the form ${}^{\Psi}g_{ij} = \Omega^2~{}^{\Phi}g_{ij}$. Then, it implies that the corresponding masses
satisfy ${}^{\Phi}m + \beta~{}^{\Psi}m \geq 0$ if ${}^{\Phi} R + \beta~\Omega^2~{}^{\Psi} R \geq 0$, for some
positive constant $\beta$. The aforementioned inequalities are fulfilled it the 
$(n-1)$-dimensional Riemannian manifolds are flat \cite{sim99}. \\
To proceed further, due to the requirement of the conformal positive energy theorem, we introduce
conformal transformations obeying the relations
\be
{}^{\Phi}g_{ij}^{\pm} = {}^{\phi}\omega_{\pm}^{2}~ \tg_{ij},
\qquad
{}^{\Psi}g_{ij}^{\pm} = {}^{\psi}\omega_{\pm}^{2}~ \tg_{ij}.
\ee
On the other hand, their conformal factors are subject to the relations
\be
{}^{\Phi}\omega_{\pm} = {\Phi_{1} \pm 1 \over 2}, \qquad
{}^{\Psi}\omega_{\pm} = {\Psi_{1} \pm 1 \over 2}.
\label{pf}
\ee
Thus, the above conformal transformations
enable one to obtain four manifolds $(\Sigma_{+}^{\Phi},~ {}^{\Phi}g_{ij}^{+})$,
$(\Sigma_{-}^{\Phi},~ {}^{\Phi}g_{ij}^{-})$, $(\Sigma_{+}^{\Psi},~ {}^{\Psi}g_{ij}^{+})$,
$(\Sigma_{-}^{\Psi},~ {}^{\Psi}g_{ij}^{+})$. 
The standard procedure of pasting 
$(\Sigma_{\pm}^{\Phi},~ {}^{\Phi}g_{ij}^{\pm})$ and 
$(\Sigma_{\pm}^{\Psi},~ {}^{\Psi}g_{ij}^{\pm})$ across the surface $V = 0$
endues to construct a regular hypersurfaces $\Sigma^{\Phi} = \Sigma_{+}^{\Phi}
\cup \Sigma_{-}^{\Phi}$
and $\Sigma^{\Psi} = \Sigma_{+}^{\Psi}
\cup \Sigma_{-}^{\Psi} $. 
If $(\Sigma,~ g_{ij},~ \Phi_{A},~ \Psi_{A})$ are asymptotically flat solution of
(\ref{ppff}) and (\ref{rr}) with non-degenerate black hole event horizon,
our next task will be to check that total
gravitational mass on hypersurfaces $\Sigma^{\Phi}$ and $\Sigma^{\Psi}$ 
is equal to zero. In order to do this we shall implement the conformal
positive mass theorem presented in Ref.\cite{sim99}. Hence, using
another conformal transformation given by
\be
\hg^{\pm}_{ij} = \bigg[ \bigg( {}^{\Phi}\omega_{\pm} \bigg)^2
 \bigg( {}^{\Psi}\omega_{\pm} \bigg)^{2 } \bigg]^{1 \over 2}\tg_{ij},
\ee
it follows that 
the Ricci curvature tensor on the space yields
\ben \label{ric}
2~\hR &=& \bigg[ {}^{\Phi}\omega_{\pm}^2~ {}^{\Psi}\omega_{\pm}^{2 \la} \bigg]
^{-{1 \over 2}}
\bigg( {}^{\Phi}\omega_{\pm}^{2} {}^{\Phi}R +
{}^{\Psi}\omega_{\pm}^{2} {}^{\Psi}R \bigg) \\ \nonumber
&+& 
\bigg( \hna _{i} \ln {}^{\Phi}\omega_{\pm} - \hna _{i} \ln {}^{\Psi}\omega_{\pm} \bigg)  
\bigg( \hna ^{i} \ln {}^{\Phi}\omega_{\pm} - \hna ^{i} \ln {}^{\Psi}\omega_{\pm} \bigg).  
\een

The close inspection of
the first term in relation (\ref{ric}) reveals that it is non-negative. Namely
one can establish that it may be written in the form as follows:
\ben
{}^{\Phi}\omega_{\pm}^{2}~ {}^{\Phi}R +
{}^{\Psi}\omega_{\pm}^{2}~ {}^{\Psi}R &=& 
2~\mid
{\Phi_{0} \tna_{i} \Phi_{-1}
- \Phi_{-1} \tna_{i} \Phi_{0} \over
\Phi_{1} \pm 1 } \mid^2 \\ \nonumber
&+&
2~\mid { \Psi_{0} \tna_{i} \Psi_{-1}
- \Psi_{-1} \tna_{i} \Psi_{0} \over
\Psi_{1} \pm 1} \mid^2.
\een
Applying the conformal energy theorem we draw a conclusion that
$(\Sigma^{\Phi},~ {}^{\Phi}g_{ij})$, $(\Sigma^{\Psi},~ {}^{\Psi}g_{ij})$ and
$(\hSi,~ \hg_{ij})$ are flat and it in turns implies that the conformal factors
${}^{\Phi}\omega = {}^{\Psi}\omega$ and $\Phi_{1} = \Psi_{1}$. Furthermore
$\Phi_{0} = const~ \Phi_{-1}$ and $\Psi_{0} = const~ \Psi_{-1}$. Just the above potentials 
are functions of a single variable. Moreover, the manifold $(\Sigma,~ g_{ij})$ is conformally flat.
We can rewrite $\hg_{ij}$ in a 
conformally flat form, i.e., we define a function
\be
\hg_{ij} = {\cal U}^{4}~ {}^{\Phi}g_{ij},
\label{gg}
\ee
where one sets ${\cal U} = ({}^{\Phi}\omega_{\pm} V)^{-1/2}$.
Because of the fact that the Ricci scalar in $\hg_{ij}$ metric is equal to zero, 
equations of motion of the system in question reduce
to the Laplace equation on the three-dimensional Euclidean manifold
\be
\na_{i}\na^{i}{\cal U} = 0,
\ee
where $\na$ is the connection on a flat manifold. 
The above equation implies that 
the following expression for the flat base space is valid. Namely, one gets
\be
{}^{\Phi}g_{ij} dx^{i}dx^{j} = \trho^{2} d{\cal U}^2 + {\tilde h}_{AB}dx^{A}dx^{B},
\ee
Then, the event horizon will be located at some constant value of ${\cal U}$.
The radius of the black hole event horizon can be terminated at fix value of
$\rho$-coordinate \cite{mar99}, which in turn can be introduced on the hypersurface $\Sigma$
by the relation $$\hg_{ij}dx^{i}dx^{j} = \rho^2 dV^2 + h_{AB}dx^{A}dx^{B}.$$
Moreover, a connected component of the event horizon can be identify at fixed value of
$\rho$.\\
Proceeding further, let us assume that
${\cal U}_{1}$ and ${\cal U}_{2}$ consist
two solutions of the boundary value problem of the system in question.
Using Green identity and integrating over the volume element, we arrive at the relation
\be
\bigg( \int_{r \rightarrow \infty} - \int_{\cal H} \bigg) 
\bigg( {\cal U}_{1} - {\cal U}_{2} \bigg) {\p \over \p r}
\bigg( {\cal U}_{1} - {\cal U}_{2} \bigg) dS = \int_{\Omega}
\mid \na \bigg( {\cal U}_{1} - {\cal U}_{2} \bigg) \mid^{2} d\Omega.
\ee
In view of the last equation, the surface integrals disappear due to the imposed boundary conditions.
On the other hand, by virtue of the above relation one finds that
the volume integral must be identically equal to zero. To summarize we have 
established the conclusion of our investigations.\\
{\it Theorem}:\\
Let us consider a static four-dimensional solution to equation of motion 
in Chern-Simons modified gravity with $U(1)$-gauge field. Suppose that one has an asymptotically
timelike Killing vector field $k_{\mu}$ orthogonal to the connected and simply
connected spacelike hypersurface $\Sigma$. The topological
boundary $\p \Sigma$ of $\Sigma$ is a nonempty topological manifold with
$g_{ij}k^{i} k^{j} = 0$ on $\p \Sigma$. It yields the following:\\
If $\p \Sigma$  is connected, then there exist a neighbourhood of the
hypersurface $\Sigma$ which is diffeomorphic to an open
set of Reissner-Nordsr\"om non-extreme solution with
{\it electric} charge.

\section{Conclusions}
In our paper we prove the uniqueness of four-dimensional static black hole being the solution
of Chern-Simons modified gravity with $U(1)$-gauge field. Assuming the existence of
an asymptotically timelike Killing vector field orthogonal to the simply connected spacelike hypersurface with 
topological boundary, it turns out that if the boundary in question is connected, then there is a neighbourhood 
of the hypersurface which is diffeomorphic to an open set
Reissner-Nordsr\"om non-extreme solution with {\it electric} charge.
\par
It may be interesting to generalize the proof to the case of both degenerate and nondegenerate 
components of the event horizon of the black hole in question. On the other hand, stationary axisymmetric
case as well as the Chern-Simons modified gravity with cosmological constant are challenges for the future investigations.
We hope to return to these problems elsewhere.


\begin{acknowledgments}
MR was partially supported by the grant of the National Science Center
$2011/01/B/ST2/00408$.
\end{acknowledgments}



\end{document}